\pdfoutput=1
\documentclass[10pt,pra,aps,superscriptaddress,nofootinbib,twocolumn,longbibliography]{revtex4-1}

\usepackage[compatibility=false]{caption} 
\usepackage{amsmath,bbm}
\DeclareMathOperator{\Tr}{Tr}
\usepackage{amsthm}
\usepackage{amsmath}
\usepackage{latexsym}
\usepackage{amssymb}
\usepackage{float}
\usepackage{graphicx}           
\usepackage{color}
\usepackage{xcolor}
\usepackage{mathpazo}
\usepackage{comment}
\usepackage{enumitem}
\usepackage{multirow}
\usepackage{setspace}
\usepackage{caption}
\usepackage{subcaption}
\usepackage{mathtools} 
\usepackage{extarrows} 
\usepackage[commandnameprefix=always]{changes}
\usepackage{svg}
\usepackage{subcaption}
\usepackage[colorlinks=true, linkcolor=blue, citecolor=blue, urlcolor=blue]{hyperref}

\usepackage{changes}

\newcommand{\be}{\begin{equation}}
\newcommand{\ee}{\end{equation}}
\newcommand{\bea}{\begin{eqnarray}}
\newcommand{\eea}{\end{eqnarray}}

\def\squareforqed{\hbox{\rlap{$\sqcap$}$\sqcup$}}
\def\qed{\ifmmode\squareforqed\else{\unskip\nobreak\hfil
\penalty50\hskip1em\null\nobreak\hfil\squareforqed
\parfillskip=0pt\finalhyphendemerits=0\endgraf}\fi}
\def\endenv{\ifmmode\;\else{\unskip\nobreak\hfil
\penalty50\hskip1em\null\nobreak\hfil\;
\parfillskip=0pt\finalhyphendemerits=0\endgraf}\fi}

\newcommand{\ket}[1]{|#1\rangle}
\newcommand{\bra}[1]{\langle#1|}

\makeatletter
\newtheorem*{rep@theorem}{\rep@title}
\newcommand{\newreptheorem}[2]{%
\newenvironment{rep#1}[1]{%
 \def\rep@title{#2 \ref{##1}}%
 \begin{rep@theorem}}%
 {\end{rep@theorem}}}
\makeatother

\newreptheorem{thm}{Theorem}

\begin{document}
\title{Quantum advantages in multiparty communication}


\author{Ankush Pandit}
\email{S26PH09009@iitbbs.ac.in\\ ankushpandit166@gmail.com}
\affiliation{School of Physics, Indian Institute of Science Education and Research Thiruvananthapuram, Kerala 695551, India}
\affiliation{Department of Physics, School of Basic Sciences, Indian Institute of Technology Bhubaneswar, Odisha 752050, India}
  
\begin{abstract}
    We investigate two senders and one receiver multiparty communication scenario. We explicitly study multiparty communication several unexplored scenario, bounded by dimension of the communication and distinguishability of senders input. We provide an explicit characterization of the classical correlations achievable under these constraints. We then demonstrate that quantum communication systematically exceeds these classical limits, even in the absence of preshared entanglement and without any input choice for the receiver. Furthermore, we also implement semidefinite hierarchy tools tailored to the two senders and one-receiver multiparty communication setting for both types of constraints considered so far. Our results reveal a clear quantum advantage in multiparty communication under those restrictions.
\end{abstract}


\maketitle

\section{Introduction}

Communication process lie at the heart of quantum information science, underpinning both its theoretical foundations and finding practical applications in various fields such as communication complexity, distributed computation, cryptography, streaming algorithm \cite{ccbook,TRbook,rao_yehudayoff_2020}. Traditionally, most of the research has been carried out one way communication task that consists one sender and one receiver \cite{PhysRevLett.105.230501, PhysRevLett.100.210503, Miklin2021universalscheme, PhysRevLett.130.080802,PhysRevA.107.062210,PhysRevLett.134.120203,glty-kkbp}, compute some function depending on their inputs, parity-oblivious multiplexing \cite{Chaturvedi2021characterising}, oblivious communication \cite{Chaturvedi2020quantum}, contextuality \cite{saha2019state,hazra2024optimaldemonstrationgeneralizedquantum,PhysRevLett.130.080802}. In one communication, the most investigation has been carried out with the constraints of bounding the dimension of the communicating systems. In another approach, instead of bounding the dimension, analogously restricting the distinguishability of the sender's inputs \cite{ Chaturvedi2020quantum, Tavakoli2022informationally}. Quantum theory significantly provides an advantage in these communication tasks and shows its superiority over any communication tasks by classical ones. \\

In this study, we are interested in characterising the correlation between senders (say Alice and Bob) and receiver (Charlie) under the restriction of both dimension and distinguishability of bounded communication. In these multiparty communication tasks, the correlations are fully characterised by set of probabilities $p(z|x,y)$ which represents, how a possibly best measurement is performed by Charlie such that his outcome $z$ depends on both of the Alice's input $x$ and Bob's input $y$. Charlie has no input choice, he has to measure with his fixed input as mentioned in \cite{PhysRevA.83.062112,pandit2025limitsclassicalcorrelationsquantum}. Bounding the communicating message dimension of both sender in dimension dimension-bounded scenario. Distinguishability on the other hand is the maximum probability of successfully guessing the sender's input from the communicating message. Analogously this also quantifies how much information about the senders input are revealed by receiver from the communicating messages.\\

We first consider the communication carried out by classical communication and characterise them in terms of facet inequalities, between Alice, Bob and Charlie. We explicitly characterise polytope as mentioned in \cite{PhysRevA.83.062112,pandit2025limitsclassicalcorrelationsquantum} and no communication are allowed from Alice to Bob or vice versa. For distinguishability  constrained communication, instead of considering a particular value of distinguishability, we characterise the polytope in terms of varying distinguishability such that for different values of distinguishability it gives a different polytope. Next we move on the quantum communication regime, where the communicating messages with quantum states and measurement performed by Charlie is POVM measurement. For the quantum communication we use the semidefinite programming (SDP) SeeSaw method to obtain the lower bound of the communication to check whether there is a quantum violation or not. We also check the SDP upper bound of the two sender and one receiver correlation similar to the \cite{PhysRevLett.98.010401} (the main difference between our hierarchy and 'NPA' hierarchy is that our settings of the game, they implement their hierarchy for  the one single preparation and spatially separated measurement, on the other hand in our case we consider two spatially preparation and one single measurement with no input choice as in \cite{Pusey_2012} type scenario). Our SDP hierarchy doesn't match with the SeeSaw bound in lower dimension but in the higher dimension, some scenario where the SeeSaw bound matches with the upper bound(NPA) considered here. We here first explicitly characterise the upper bound in a multiparty communication both in dimension and distinguishability bounded scenario.

\section{Preliminaries}
\begin{figure}[h]
    \centering
    \includegraphics[width=0.8\linewidth]{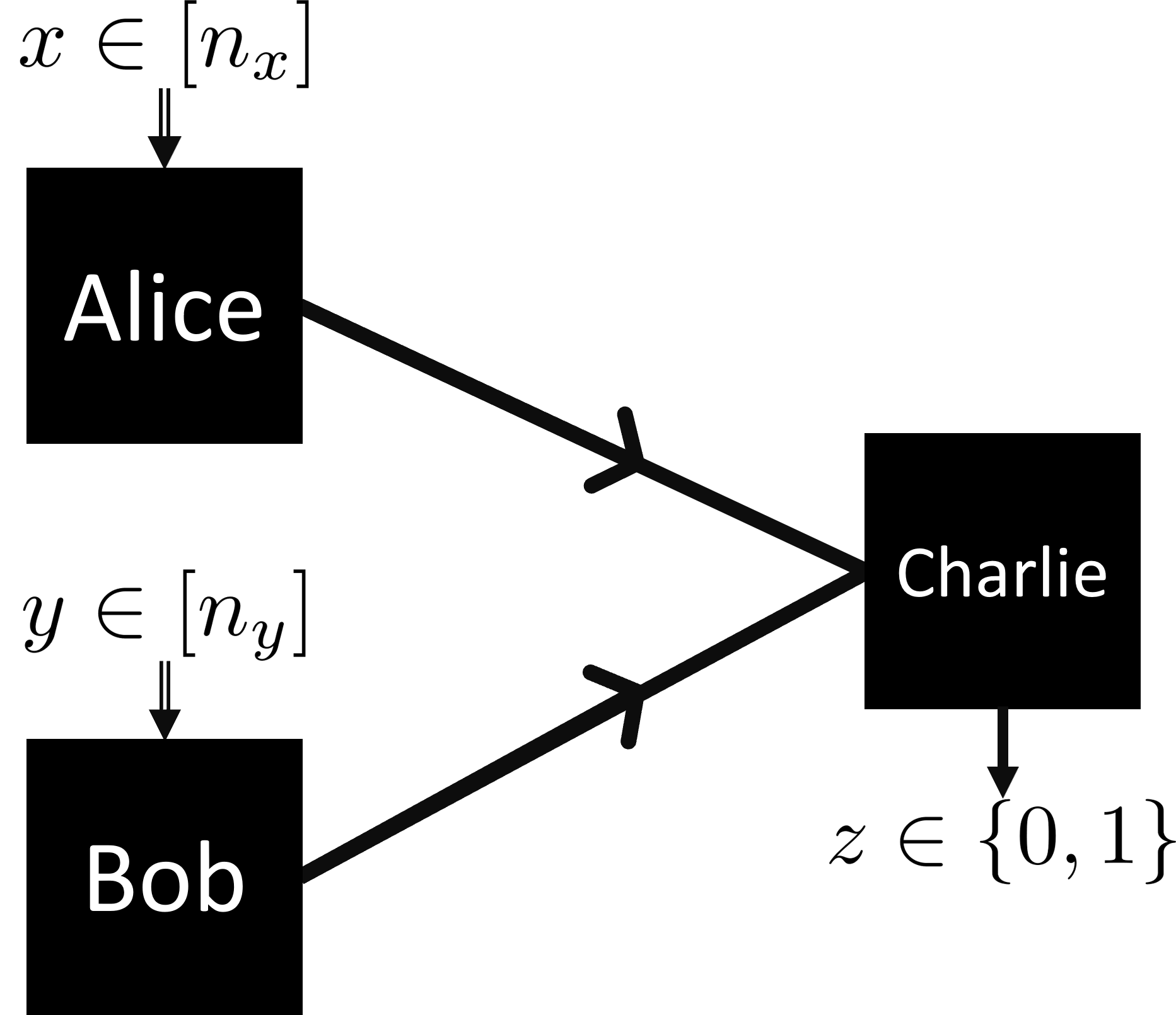}
    \caption{Schematic diagram of multiparty communication. Where spatially separated senders are represented by Alice and Bob and receiver is Charlie with no input.}
    \label{fig1}
\end{figure}

Let us consider a communication scenario consists of two spatially separated senders let's say Alice and Bob, and receiver Charlie as shown in the Fig \ref{fig1}. Both Alice and Bob receives an inputs $x \in  [n_x]$ and $y \in  [n_y]$ (here $[N]$ represents the set $\{1,\cdots,N \}$) respectively, and separately communicates with Charlie, i.e, no communication between Alice to Bob or vice versa are allowed. Charlie has no inputs or equivalently one can consider as a fixed input choice. Depending on the messages from Alice and Bob, Charlie performs some measurement and obtains an outcome $z\in [n_z]$. We denotes this task as $(n_x,n_y,n_z)$ (where $n_x$ is number of Alice's inputs, $n_y$ is number of Bob's inputs and $n_z$ is number of Charlie's outcomes) respectively. In each round of the game, we characterize the task with a probability $p(z|x,y)$, which represents the probability of obtaining an outcome $z$, is Alice's input $x$ and Bob's input $y$ and Charlie gets an outcome $z$. The constraints of the communication task we consider here in two ways,\\
\\

(1). \textbf{Dimension Bound:} In multiparty quantum communication bounded by communicating message dimension, was first introduced by \cite{PhysRevA.83.062112}. Here we study several unexplored scenarios about multiparty communication for different inputs and outputs from their settings. This task can be done if we restrict the allowed communication between Alice, Bob and Charlie is bounded by two. If Alice and Bob depending on their input $x$ and $y$, communicates with Charlie by sending messages $m \in \{1,2\}$ and $n \in \{1,2\}$ respectively, with probability $p(m|x)$ and $p(n|y)$ respectively. Charlie's measurement based on the messages $m ~\text{and} ~n$, obtains an outcome $z$ with probability $p(z|m,n)$. The observed probability arises between Alice, Bob and Charlie is given by
\bea
p(z|x,y) &=& \sum_{m,n} p(m|x)p(n|y)p(z|m,n)
\eea
The figure of merit of the performance of this task can be written as $\mathcal{S_C}=c(x,y,z) p(z|x,y)$. \\

(2). \textbf{Distinguishability Bound:}  This task can be thought as Alice and Bob don't want to reveal about the distinguishability of their inputs $x$ and $y$, rather they bound their input's distinguishability with some value $D_1$(for Alice) and $D_2$(for Bob) and send messages $m$ and $n$ to Charlie, with probability $p_e(m|x)$ and $p_e(n|y)$ respectively. In general the number of messages for each sender is $m=2^{n_x-1}$, $n=2^{n_y-1}$, where $n_x,n_y$ is number of inputs of Alice and Bob sides, as discussed in \cite{pandit2025limitsclassicalcorrelationsquantum}. Mathematically, we can write it as
\bea
    \sum_m \max_{x}  q_x p_e(m|x) \leqslant D_1, \nonumber \\ \label{eq2}
\sum_n \max_{y} q_y p_e(n|y) \leqslant D_2.
\eea
Where $q_i, i\in \{x,y\}, q_i \geqslant0, \sum_i q_i=1, \forall i$ is the apriori probability distribution of choosing an input among all total inputs they receive. Charlie, depending on these messages performs measurement and obtains as outcome $z$ with probability $p_d(z|m,n)$, which we refer as Charlie's decoding strategy. The observed probability arises between Alice, Bob and Charlie is given by,
\bea
p(z|x,y)=\sum_{m,n}p_d(z|m,n)p_e(m|x)p_e(n|y). \\
\eea
with $p_e(m|x)$ and $p_e(n|y)$ is bounded by some value as given in eq (\ref{eq2}). The figure of merit of the performance of this task can be written as $\mathcal{S_C}=c(x,y,z) p(z|x,y)$. \\

\section{Quantum Communication}
\subsection{Dimension Bound}
In multiparty quantum communication, for each input $x$ of Alice and $y$ of Bob, sends qubit states $\rho_x$ and $\sigma_y$ to Charlie. Charlie in turn depending on these states, performs POVM measurements $M_z \geqslant 0, \sum_z M_z = \mathbb{I}$. The observed probabilities arises between Alice, Bob and Charlie in quantum regime is given by,
\bea
p(z|x,y) = \Tr [(\rho_x \otimes \sigma_y) M_z]. \\
\eea
The figure of merit of the performance of this task can be written as $\mathcal{S_Q}=c(x,y,z) p(z|x,y)$. \\
\subsection{Distinguishability Bound}
In quantum communication under distinguishability bound, Alice and Bob, sends any qudit states $\rho_x$ and $\sigma_y$ (they can send any dimension states because there is no bound in dimension in this scenario) to Charlie. Both $\rho_x$ and $\sigma_y$ also bounded their distinguishability. Let $\Theta$ and $\Phi$ are two auxiliary variables such that $\Theta \geqslant \rho_x, q_x \Theta \leqslant D_1$ and $\Phi \geqslant \sigma_y; q_y\Phi \leqslant D_2$. Charlie according the states performs POVM measurements $M_z \geqslant 0, \sum_z M_z =\mathbb{I}$. The observed probabilities arises between Alice, Bob and Charlie in quantum regime is given by,
\bea
p(z|x,y) = \Tr [(\rho_x \otimes \sigma_y) M_z].
\eea
The figure of merit of the performance of this task can be written as $\mathcal{S_Q}=c(x,y,z) p(z|x,y)$.
\subsection{Quantifying advantage}
In order to show quantum advantage over classical, we define the advantage $\mathcal{S_Q} > \mathcal{S_C}$ for any communication scenario discussed so far. There is no quantum advantage in any scenario if $\mathcal{S_Q} \leqslant \mathcal{S_C}$. We obtain the maximum classical $\mathcal{S_C}$ by using linear programming and using See-Saw optimization technique we obtain the lower bound for quantum $\mathcal{S_Q}^L_d$ (superscript $L$ stands for lower bound and subscript $d$ stands for dimension of the communicating systems). We also provide semidefinite hierarchy to get the upper bound in multiparty communication, for both dimension bound and distinguishability bound, respectively. For dimension bound multiparty communication scenario  $\mathcal{S_Q}^L_d$ can be obtained from SeeSaw optimization as discussed in \cite{PhysRevA.83.062112}. For distinguishability bounded multiparty communication scenario $\mathcal{S_Q}^L_d$ also obtained by SeeSaw discussed in \cite{pandit2025limitsclassicalcorrelationsquantum}. For semidefinite hierarchy bound $\mathcal{S_Q}$ discussed in Appendix \ref{hierarchy}. In this study \texttt{Python} and \texttt{MATLAB} are used to generate the extremal points of polytope of $(n_x,n_y,n_z)$ scenarios and \texttt{PANDA} is used to generate the facets of the polytope \cite{LORWALD2015297}.

\section{Results}
\subsection{(3,2,2) dimension bound scenario}
In this scenario we enlist the non-trivial facet inequalities in Table \ref{table1}. Let us consider $\mathcal{I}_1=-p(1|1,2) -p(1|2,1) +p(1|2,2) +p(1|3,1) +p(1|3,2)$. The maximum classical bound for $\mathcal{I}$ is $\mathcal{S_C}=2$, wheareas using See-Saw optimization we get $\mathcal{S_Q}^L_2=2.4142$. This investigation clearly suggests $\mathcal{S_Q}^L_2 > \mathcal{S_C}$ as an advantage of quantum communication over classical ones. \\

\textit{An explicit qubit strategy for getting this quantum violation of $\mathcal{I}_1$}: Let us consider Alice prepares states on her side $\ket{\psi_1} = \ket{0}$, $\ket{\psi_2}=\sin{\left(\frac{\pi}{8}\right)}\ket{0} + \cos{\left(\frac{\pi}{8}\right)}\ket{1}$ and $\ket{\psi_3} = -\sin{\left(\frac{\pi}{8}\right)}\ket{0} + \cos{\left(\frac{\pi}{8}\right)}\ket{1}$. Bob also prepare states in his side given as $\ket{\phi_1}=\ket{0}$ and $\ket{\phi_2}=\ket{1}$. The Charlie measurement is given by $M_1=\ket{\eta_1}\bra{\eta_1}+\ket{\eta_2}\bra{\eta_2}$, where $\ket{\eta_1} = \left( 0,0,0,1 \right)^T$, $\ket{\eta_2} = \left( \frac{1}{\sqrt{2}},0,\frac{1}{\sqrt{2}},0 \right)^T$ and $M_0=\mathbb{I} - M_1$. Using these states and measurement on can achieve $\mathcal{S_Q}_2^L = 2.4142$.\\

\subsection{(4,2,2) dimension bound scenario}
In this scenario we enlist the non-trivial facet inequalities in Table \ref{table2}. Consider for the inequality $\mathcal{I}_2=p(1|1,1) +p(1|1,2) +p(1|2,1) -p(1|2,2) -p(1|3,1) +p(1|3,2) -p(1|4,1) -p(1|4,2)$. The maximum classical value for $\mathcal{I}_2$, is given as $\mathcal{S_C}=2$. But using See-Saw we get $\mathcal{S_Q}^L_2 = 2.8284$. This investigation clearly suggests $\mathcal{S_Q}^L_2 > \mathcal{S_C}$ as a advantageous of quantum communication over classical ones. \\

\textit{An explicit qubit strategy for getting this quantum violation of $\mathcal{I}_2$}: Let us consider Alice prepares states $\ket{\psi_1} = \ket{0}$, $\ket{\psi_2}=\ket{+}$, $\ket{\psi_3}=\ket{-}$ and $\ket{\psi_4}=\ket{1}$. And Bob also prepares states $\ket{\phi_1}=\ket{0}$ and $\ket{\phi_2}=\ket{1}$ respectively. Charlie's measurement is given by $M_1=\ket{\eta_1}\bra{\eta_1}+\ket{\eta_2}\bra{\eta_2}$, where $\ket{\eta_1} = \left( 0,-\cos{\left(\frac{\pi}{8}\right)},0,\sin{\left(\frac{\pi}{8}\right)} \right)^T$, $\ket{\eta_2} = \left(\cos{\left(\frac{\pi}{8}\right)},0,\sin{\left(\frac{\pi}{8}\right)},0 \right)^T$ and $M_0=\mathbb{I} - M_1$. Using these states and measurement on can achieve $\mathcal{S_Q}_2^L = 2.8284$.\\ 

\subsection{(3,2,3) dimension bound scenario}
In this scenario we enlist the facet inequalities in Table \ref{table323}. Consider a inequality $\mathcal{I}_3=2p(1|1,1) +p(1|1,2) -p(1|2,2) -2p(1|3,1) +p(1|3,2) +p(2|1,1) +p(2|2,1) -p(2|3,1)$.The maximum classical bound is $\mathcal{S_C}=3$, but using SeeSaw we get $\mathcal{S_Q}_2^L = 3.25$. This also suggests $\mathcal{S_Q}^L_2 > \mathcal{S_C}$ as a advantageous of quantum communication over classical ones. \\

\textit{An explicit qubit strategy for getting this quantum violation of $\mathcal{I}_3$}: Consider Alice prepares three states $\ket{\psi_1}=\ket{0}$ $\ket{\psi_2}=\cos{\left(\frac{\pi}{3}\right)} \ket{0} + \sin{\left(\frac{\pi}{3}\right)} \ket{1}$, 
$\ket{\psi_3}= - \cos{\left(\frac{\pi}{3}\right)} \ket{0} + \sin{\left(\frac{\pi}{3}\right)} \ket{1}$. Similarly Bob also prepares two states in his side as given as $\ket{\phi_1}=\ket{0}$ and $\ket{\phi_2}=\ket{1}$. Charlie's POVM are given by, $M_1=\frac{7}{8} \ket{\eta_1}\bra{\eta_1}+\ket{\eta_2}\bra{\eta_2}$, $M_2=\frac{1}{4} \ket{\delta_1}\bra{\delta_1}+\frac{1}{4} \ket{\delta_2}\bra{\delta_2}$, and $M_0=\mathbb{I}-M_1-M_2$, where $\ket{\eta_1}=\left(- \sqrt{\frac{27}{28}},0, - \sqrt{\frac{1}{28}}, 0 \right)^T$, $\ket{\eta_2}=\left(0, - \frac{\sqrt{3}}{2}, 0, \frac{1}{2} \right)^T$, $\ket{\delta_1}=\left( \frac{1}{2},0, \frac{\sqrt{3}}{2}, 0 \right)^T$, $\ket{\delta_2}=\left(0, \frac{1}{2}, 0, \frac{\sqrt{3}}{2} \right)^T$. Using these states and measurements we get $\mathcal{S_Q}_2^L = 3.25$.\\ 

\subsection{(4,3,2) dimension bound scenario}
In this scenario we enlist the non-trivial facet inequalities in Table \ref{table3}. Consider for the inequality $\mathcal{I}_4=9p(1|1,1) -4p(1|1,2) -9p(1|1,3) +p(1|2,1) +4p(1|2,2) +5p(1|2,3) -5p(1|3,1) -2p(1|3,2) +3p(1|3,3) -9p(1|4,1) +6p(1|4,2) -15p(1|4,3)$. The maximum classical value for $\mathcal{I}_3$, is given as $\mathcal{S_C}=10$. But using See-Saw we get $\mathcal{S_Q}^L_2 = 13.3843$. This investigation clearly suggests $\mathcal{S_Q}^L_2 > \mathcal{S_C}$ as an advantage of quantum communication over classical ones.\\

Consider again for another inequality in this scenario $\mathcal{I}_5=5p(1|1,1) + 3p(1|1,2) - 2p(1|1,3) + 2p(1|2,1) - 5p(1|2,2) + 3p(1|2,3) - 5p(1|3,1) + 2p(1|3,2)+5p(1|3,3) - 6p(1|4,1) - 2p(1|4,2) - 8p(1|4,3)$. The maximum classical value for $\mathcal{I}_3$, is given as $\mathcal{S_C}=8$. But using See-Saw, we get $\mathcal{S_Q}^L_2 = 10.8596$. This investigation also suggests $\mathcal{S_Q}^L_2 > \mathcal{S_C}$ as an advantage of quantum communication over classical ones.\\

\textit{An explicit qubit strategy for getting this quantum violation of $\mathcal{I}_5$}: Let us consider Alice prepares four states on her side $\ket{\psi_1}=\ket{0}$, $\ket{\psi_2}=-\sin{\left(\frac{\pi}{13}\right)}\ket{0} + \cos{\left(\frac{\pi}{13} \right)} \ket{1}$,  $\ket{\psi_3}=-\sin{\left(\frac{\pi}{9}\right)}\ket{0} + \cos{\left(\frac{\pi}{9}\right)} \ket{1}$, $\ket{\psi_4}=\sin{\left(\frac{\pi}{6}\right)}\ket{0} + \cos{\left(\frac{\pi}{6}\right)} \ket{1}$. Similarly Bob also prepare three states in his side given as $\ket{\phi_1}=\ket{0}$, $\ket{\phi_2}=e^{i\pi}\ket{0}$, $\ket{\phi_3}=\ket{1}$. Charlie performs POVM measurements as given by $M_1=\ket{\eta_1}\bra{\eta_1}+\ket{\eta_2}\bra{\eta_2}$, where $\ket{\eta_1} = \left(\sin{\left(\frac{\pi}{30}\right)},0,\cos{\left(\frac{\pi}{30}\right)},0 \right)^T$, $\ket{\eta_2} = \left(0, \sin{\left(\frac{\pi}{5}\right)},0,\sin{\left(\frac{\pi}{5}\right)} \right)^T$ and $M_0=\mathbb{I} - M_1$. Using these states and measurement on can achieve $\mathcal{S_Q}_2^L \approx 10.85$.\\

\subsection{(3,3,2) distinguishability bound scenario}
In this scenario, bounded by the distinguishability of Alice's and Bob's inputs, we enlist the facet inequalities that falls under this scenario and their quantum violations in Table \ref{table4}. We listed the $\mathcal{S_Q}_d^L$ upto $d=2$, and $\mathcal{S_Q}$ that we get from the semidefinte hierarchy discussed in Section \ref{hierarchy}.

Let us consider the inequality ~$\mathcal{I}_6= p(1|1,1) -p(1|2,1) -p(1|3,1) -2p(1|1,2) +2p(1|2,2) +p(1|3,2) +p(1|1,3) +3p(1|2,3) -2p(1|3,3) \leqslant 6D1 +3D2 -1$. For $D_1=D_2=2/3$, we get $\mathcal{S_C}=5$, whereas by using See-Saw optimization we get $\mathcal{S_Q}^L_2=5.5348$. This investigation clearly suggests quantum violation $\mathcal{S_Q}^L_2 > S_C$ as an advantage of quantum communication over classical ones.

\textit{An explicit qubit strategy for getting this quantum violation of $\mathcal{I}_6$}: Consider Alice prepares three qubit states in her side are given as $\ket{\psi_1}=\ket{0}$, $\ket{\psi_2}=\ket{+}$, $\ket{\psi_3}= -\sin{\left(\frac{4\pi}{89}\right)}\ket{0} + \cos{\left(\frac{4\pi}{89}\right)}\ket{1} $. Similarly Bob also prepares qubit states in his side are given as $\ket{\phi_1}=\ket{0}$, $\ket{\phi_2}=\ket{1}$ and $\ket{\phi_3}=\ket{0}$. Charlie performs POVM measurements given as $M_1=\ket{\eta_1}\bra{\eta_1}+\ket{\eta_2}\bra{\eta_2}$, where $\ket{\eta_1}= \left( \sqrt{\frac{69}{74}},0, \sqrt{\frac{15}{69}},0 \right)^T $, $\ket{\eta_2}= \left( 0, \sqrt{\frac{15}{69}},0, \sqrt{\frac{69}{74}} \right)^T $. Using these states and measurement on can achieve $\mathcal{S_Q}_2^L = 5.5348$.\\
\subsection{Semidefinte hierarchy } \label{sdphierarchy}
We check upper bound for multiparty communication bounded by both dimension and distinguishability. To date most of the hierarchy has been studied in these two constraints scenario based on the prepare and measure scenario, but less focused on implementing the hierarchy on the multiparty communication. For spatially separated measurements it is already known that the 'NPA'-hierarchy method for upper bounding the correlation. We here study for the case of spatially separated sender and one single receiver where receiver has no input but still we get the quantum violation. \\

Let us consider the inequality $\mathcal{I}_1 \leqslant 2$.  The $\mathcal{S_Q}_2^L=2.4142$, whereas the our hierarchy gives the bound $\mathcal{S_Q}=3$. In $d=2$, our hierarchy doesn't match with the lower bound. We test in some other dimension $d$ where it converges with the SeeSaw lower bound. If we consider the case $d=3$, we get $\mathcal{S_Q}_3^L=\mathcal{S_Q}=3$. \\

\begin{figure}[H]
    \centering
    \includegraphics[width=1.11\linewidth]{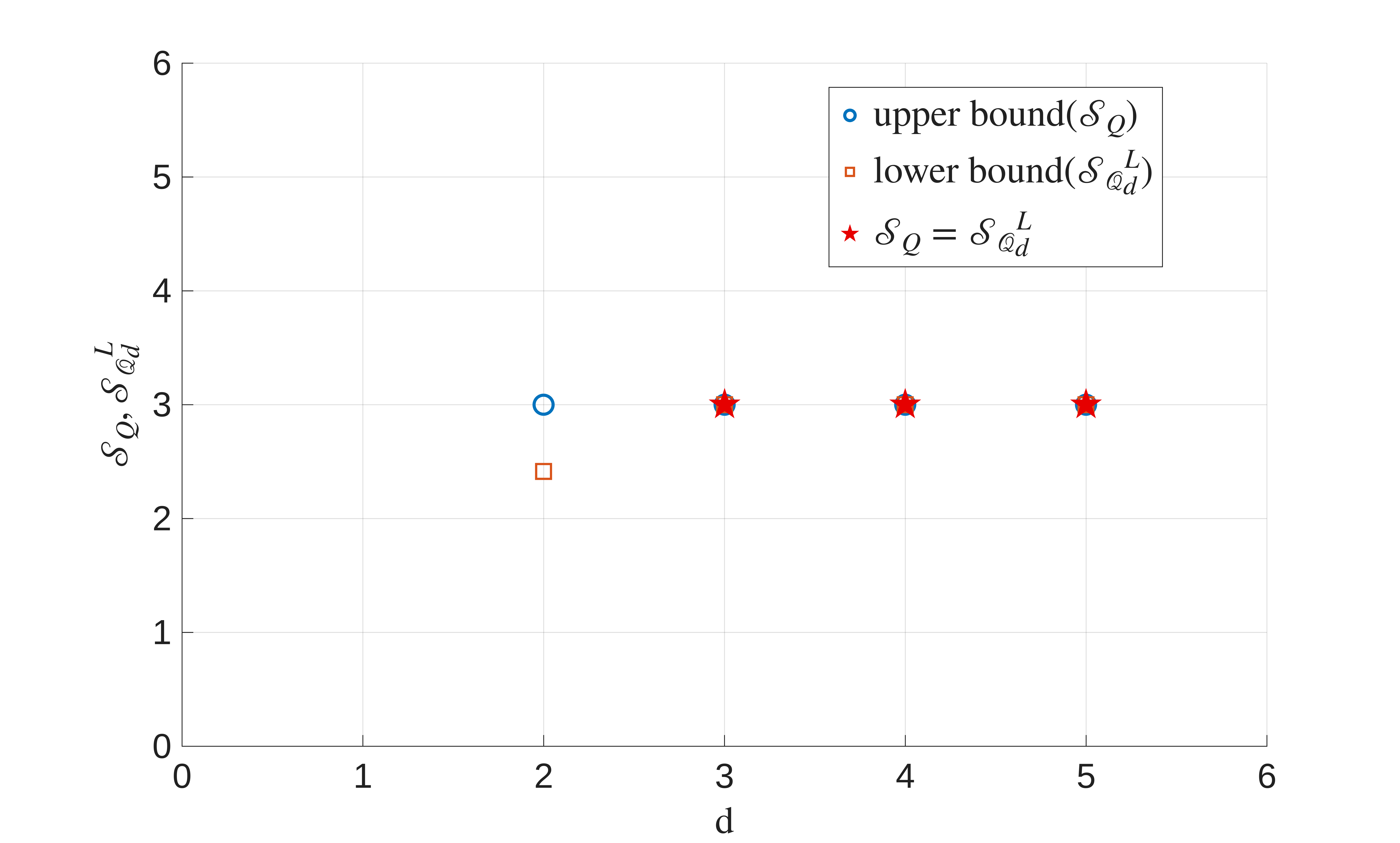}
    \caption{$\mathcal{S_Q}$ and $\mathcal{S_Q}_d^L$ for $d=2,3,4,5$ of the inequality $\mathcal{I}_1$.}
    \label{fig:placeholder}
\end{figure}

\begin{figure}[H]
    \centering
    \includegraphics[width=1.11\linewidth]{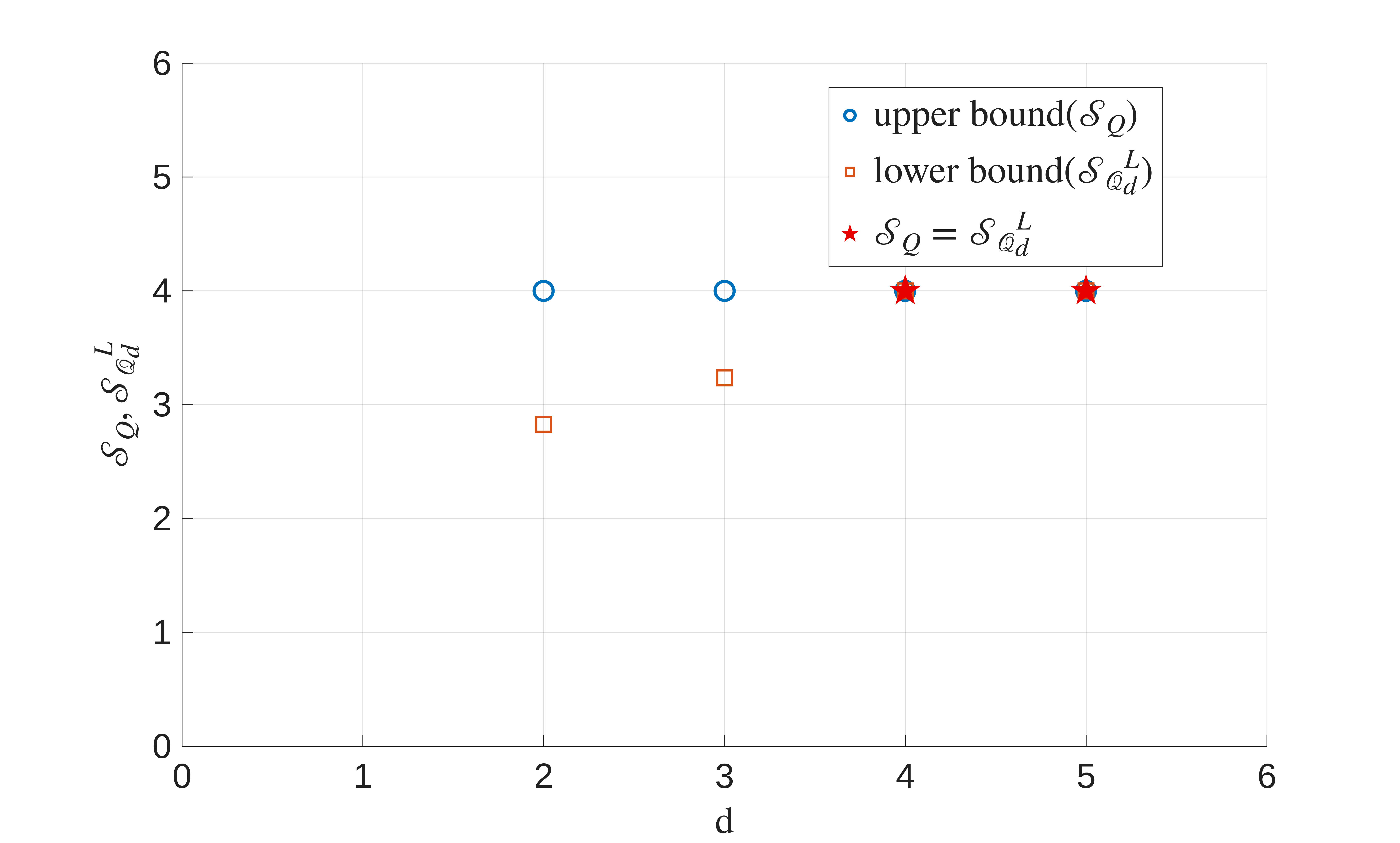}
    \caption{$\mathcal{S_Q}$ and $\mathcal{S_Q}_d^L$ for $d=2,3,4,5$ of the inequality $\mathcal{I}_2$.}
    \label{422_fig}
\end{figure}

Similarly one can consider $\mathcal{I}_2$, whose $\mathcal{S_Q}_2^L= 2\sqrt{2}$, but $\mathcal{S_Q}=4$. The comparison of lower and upper bound as depicted in the Fig \ref{422_fig}.

\begin{figure}[H]
    \centering
    \includegraphics[width=1.11\linewidth]{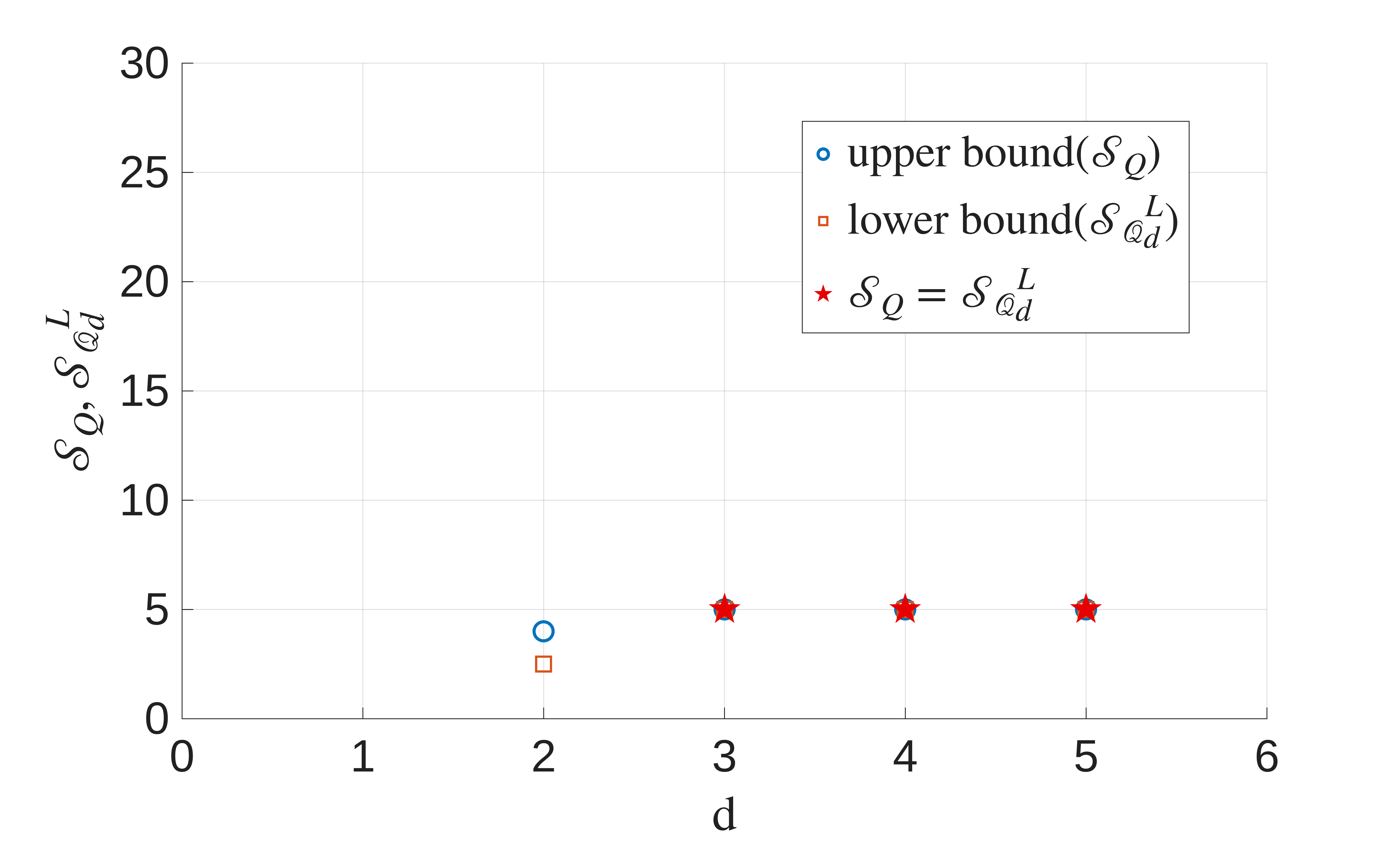}
    \caption{$\mathcal{S_Q}$ and $\mathcal{S_Q}_d^L$ for $d=2,3,4,5$ of the inequality $\mathcal{I}_3$.}
    \label{323fig}
\end{figure}

For the inequality $\mathcal{I}_3$ we depicted the comparison between $\mathcal{S_Q}$ and $\mathcal{S_Q}_d^L$ for $d=2,3,4,5$ as shown in Fig \ref{323fig}\\

Again if we consider for the inequality $\mathcal{I}_4$ in (4,3,2)- dimension bound scenario the upper bound we get $\mathcal{S_Q}=16$ for $d=2$,  and for $d=3$, $\mathcal{S_Q}=20$ but $\mathcal{S_Q}_3^L= 15.4286$. This will match for $d=4$ in SeeSaw $\mathcal{S_Q}_4^L = \mathcal{S_Q}= 20$ ($\mathcal{S_Q}$ is calculated for $d=3$).\\
\begin{figure}[H]
\hspace{-0.8cm}
    \centering
    \includegraphics[width=1.11\linewidth]{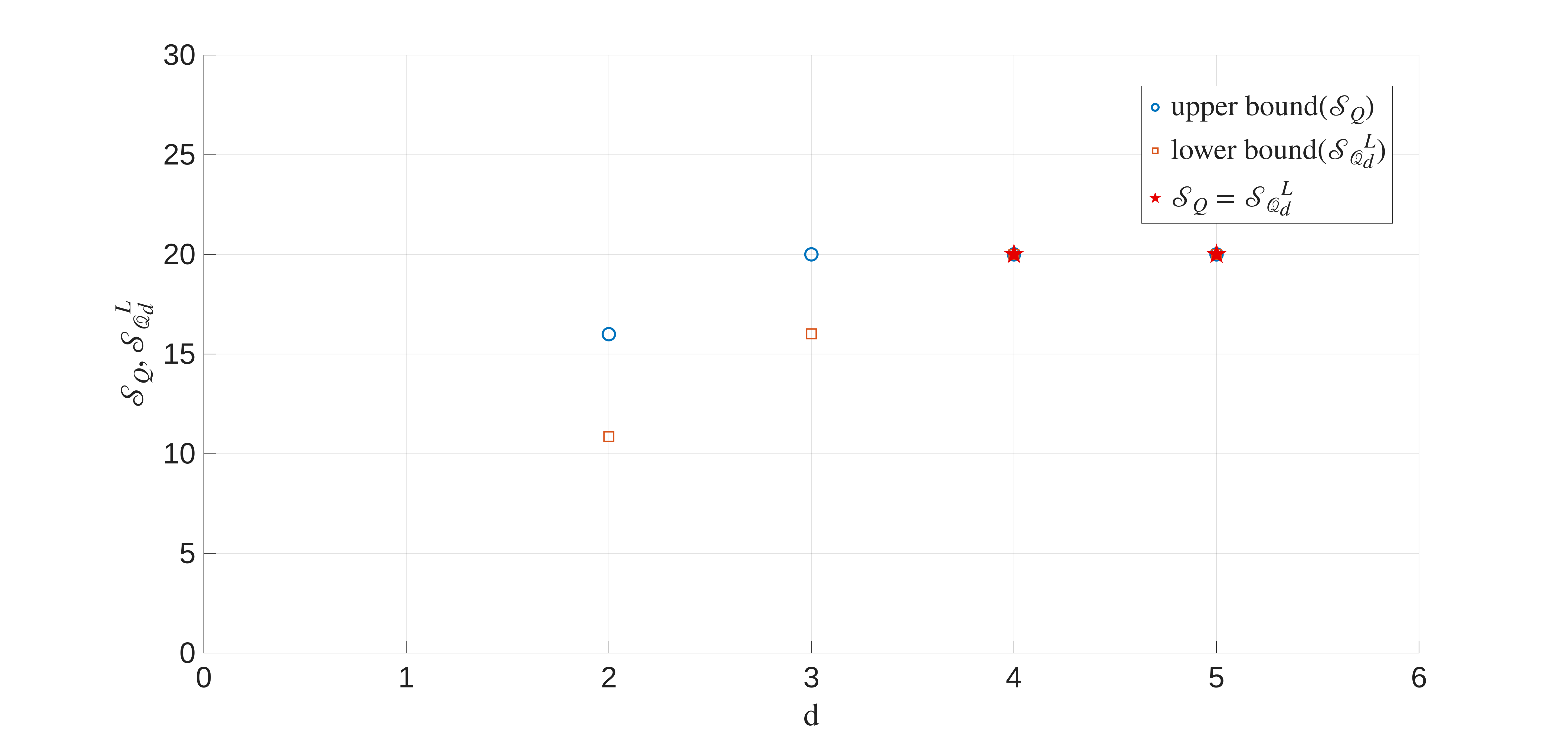}
    \caption{$\mathcal{S_Q}$ and $\mathcal{S_Q}_d^L$ for $d=2,3,4,5$ of the inequality $\mathcal{I}_4$.}
    \label{fig:placeholder}
\end{figure}

\begin{figure}[H]
    \centering
    \includegraphics[width=1.11\linewidth]{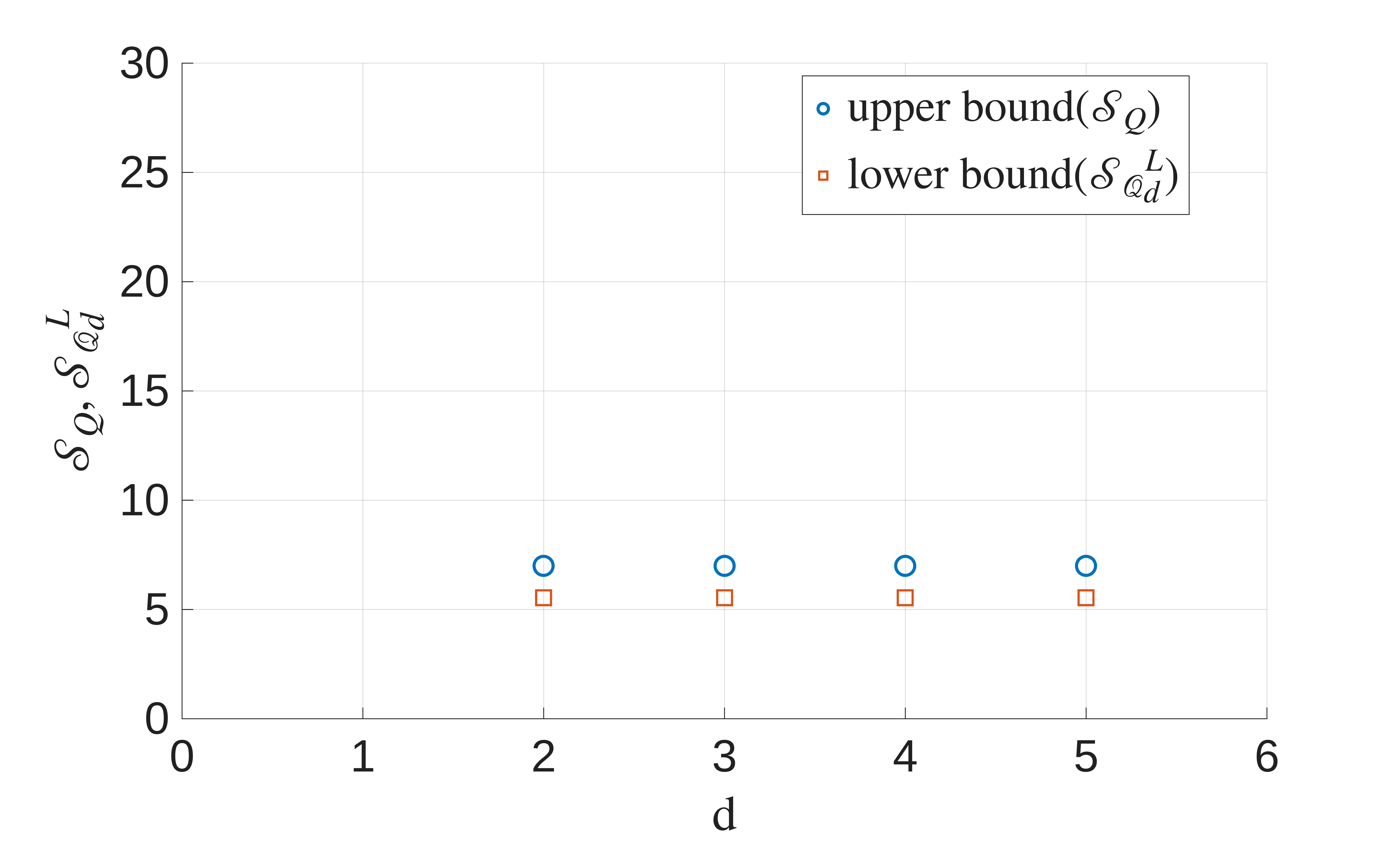}
    \caption{$\mathcal{S_Q}$ and $\mathcal{S_Q}_d^L$ for $d=2,3,4,5$ of the inequality $\mathcal{I}_6$.}
    \label{332figdist}
\end{figure}

Consider of inequality $\mathcal{I}_6$ in (3,3,2) distinguishability bounded scenario. In this case we checked up to dimension $d=5$, we don't find any points where upper and lower bound matches. There might be some other high dimension $d$, where this two bound matches. The comparison of upper and lower bound as depicted in the Fig \ref{332figdist}.\\

\section{Conclusion}
In this study, we consider two sender and one receiver multiparty communication scenario bounded by the dimension of their communicating message and distinguishability of the senders inputs. We explicitly listed the facet inequalities and their corresponding quantum violation, and show that these violations clearly indicates the quantum superiority over the classical ones in this two constrained communication scenarios. We also enlists the upper bound by using the semidefinite hierarchy in these two cases. As of our knowledge no one did this before for two spatially separated sender and one receiver communication task. Although our hierarchy did not converge with the SeeSaw method, but in higher dimensional SeeSaw value converge with the upper bound.\\

(1) In future one may interested to implement to get the tighter bound in semidefinte hierarchy in a multiparty communication. (2) One can extend this analysis to more general communication networks can reveal more interesting features of quantum communication. (3) One can increase inputs, outputs or increase the number of party in the communication games and check the how good quantum violations are there and the results generalize to $N$ party. (4) One may study the role of \textit{}shared entanglement and investigate whether there will be more quantum advantage or not.\\

\section{Acknowledgement}
The author thanks to Debashis Saha, Anubhav Chaturvedi and Satyaki Manna for fruitful discussions. The author also thanks UGC, India for the Junior Research Fellowship (reference no.: 221610057949).
\section{Statement on AI}
No AI tools are used in this study. The results are due to the author, who take full responsibility for the contents in the manuscript.
\bibliographystyle{unsrtnat}
\bibliography{cite}

\appendix

\onecolumngrid

\section{Dimension bound scenario}

\subsection{($3,2,2$) scenario} \label{322scenario}
In this scenario we get total $512$ extremal points from these we get facet inequalities. All of them does not produce any quantum advantage, and some of them are trivial inequalities. Here we enlist in Table \ref{table1}, only those inequalities that have quantum quantum violation $\mathcal{S_Q}_d^L$ up to dimension $d=2$ and $\mathcal{S_{Q}}_2^{max}$ corresponding the facet.
\begin{table}[htpb] 
    \centering
    \begin{tabular}{|c|c|c|c|c|}
    \hline
    Inequalities &  $\mathcal{S_Q}_2^L$&  $\mathcal{S_{Q}}_2^{max}$  \\ \hline
       $\mathcal{I}_1= -p(1|1,2) -p(1|2,1) +p(1|22) +p(1|3,1) +p(1|3,2) \leqslant 2 $  &$2.4142$  &$3$ \\ \hline
        $ -p(1|1,1) +p(1|1,2) +p(1|2,1) -p(1|3,1) -p(1|3,2) \leqslant 1$. &$1.4142$ & 2 \\ \hline
    \end{tabular}
    \caption{Facets of $(3,2,2)$ dimension bound scenario and their quantum violations are enlisted. $\mathcal{S_Q}_2^{max}$ denotes npa hierarchy bound.}
    \label{table1}
\end{table}

\subsection{$(4,2,2)$ scenario}
In this scenario we get total $1024$ extremal points from these we get facet inequalities. All of them does not produce any quantum advantage, and some of them are trivial inequalities. Here we enlist in Table \ref{table2}, only those inequalities that have quantum violation. \\
\begin{table}[htpb]
    \centering
    \begin{tabular}{|c|c|c|c|c|}
    \hline
    Inequalities &  $\mathcal{S_Q}_2^L$& $\mathcal{S_{Q}}_2^{max}$  \\ \hline
      $p(1|1,1) -p(1|3,1) +p(1|3,2) -p(1|4,1) -p(1|4,2) \leqslant 1$   &1.4142  & 2 \\ \hline
      $\mathcal{I}_2=p(1|1,1) +p(1|1,2) +p(1|2,1) -p(1|2,2) -p(1|3,1) +p(1|3,2) -p(1|4,1) -p(1|42) \leqslant 2$ ~~& 2.8284 & 4 \\ \hline

      $p(1|1,1) +p(1|1,2) +p(1|2,1) -p(1|2,2) -p(1|4,1) \leqslant 2 $ & 2.4142& 3\\ \hline
      
    \end{tabular}
    \caption{Facets of $(4,2,2)$ dimension bound scenario and their quantum violations are enlisted. $\mathcal{S_Q}_2^{max}$ denotes npa hierarchy bound.}
    \label{table2}
\end{table}\\

\subsection{(3,2,3) scenario}

In this scenario we get $2592$ number of extremal points, and using these points we get the facet inequalities. We enlisted those inequalities in Table \ref{table323}.

\begin{table}[ht]
    \centering
    \begin{tabular}{|c|c|}
    \hline
    Inequalities &  $\mathcal{S_Q}_2^L$  \\ \hline

      $p(1|1,1) +p(1|1,2) +p(1|2,1) -p(1|2,2) -p(1|3,1) \leqslant 2$ ~&~ 2.4142 \\ \hline

      $p(1|1,1) +p(1|1,2) +p(1|2,1) -p(1|2,2) -p(1|3,1) +p(2|1,2) -p(2|2,2) \leqslant 2 $ ~&~ 2.4142  \\ \hline

      $2p(1|1,1) -2p(1|3,1) +p(2|1,1) -p(2|1,2) +p(2|2,1) +p(2|2,2) -p(2|3,1) -p(2|3,2) \leqslant 2$ ~&~ 2.25 \\ \hline
$2p(1|1,1) -2p(1|3,1) +p(2|1,1) +p(2|1,2) +p(2|2,1) -p(2|2,2) -p(2|3,1) +p(2|3,2) \leqslant 3$ ~&~ 3.25 \\ \hline
$\mathcal{I}_3=2p(1|1,1) +p(1|1,2) -p(1|2,2) -2p(1|3,1) +p(1|3,2) +p(2|1,1) +p(2|2,1) -p(2|3,1) \leqslant 3$ ~&~ 3.25 \\ \hline

$2p(1|1,1) +p(1|1,2) -p(1|2,2) -2p(1|3,1) +p(1|3,2) +p(2|1,1) +p(2|1,2) +p(2|2,1)$ ~&~ \\ 
$-p(2|2,2) -p(2|3,1) +p(2|3,2) \leqslant 3$ ~&~ 3.2361\\ \hline

    \end{tabular}
    \caption{Facets of $(3,2,3)$ dimension bound scenario and their quantum violations are enlisted.}
    \label{table323}
\end{table}
\subsection{(4,3,2) scenario}

In this scenario we get total $2048$ number of extremal points, and from these we get the facet inequalities. Some of the facets we get are trivial, we enlist some facets in Table \ref{table3}, and their quantum violation $\mathcal{S_Q}_d^L$ up to dimension $d=2$ corresponding the facet. 

\begin{table}[htpb]
    \centering
    \begin{tabular}{|c|c|}
    \hline
    Inequalities &  $\mathcal{S_Q}_2^L$  \\ \hline
      $\mathcal{I}_4=9p(1|1,1) -4p(1|1,2) -9p(1|1,3) +p(1|2,1) +4p(1|2,2) +5p(1|2,3) -5p(1|3,1) -2p(1|3,2) +3p(1|3,3) $ &\\$ -9p(1|4,1) +6p(1|4,2) -15p(1|4,3) \leqslant 10$   & 13.3843 \\ \hline

      $7p(1|1,1) -7p(1|1,3) +p(1|2,1) -3p(1|2,2) -p(1|2,3) -2p(1|3,1) +3p(1|3,2) +5p(1|3,3) -8p(1|4,1) $&\\$ +5P(1|4,2) -13p(1|4,3) \leqslant 8$ & 10.4865  \\ \hline

      $8p(1|1,1) +5p(1|1,2) -3p(1|1,3) +6p(1|2,1) -10p(1|2,2) -6p(1|2,3) -2p(1|3,1) +2p(1|3,2) $ &\\$+4p(1|3,3) -6p(1|4,1) +7p(1|4,2) -13p(1|4,3) \leqslant 14$ & 16.3930  \\ \hline

      $11p(1|1,1) +p(1|1,2) -12p(1|1,3) +3p(1|2,1) -3p(1|2,2) +6p(1|2,3) -3p(1|3,1) +5p(1|3,2) $&\\$ +6p(1|3,3) -11p(1|4,1) -p(1|4,2) -10p(1|4,3) \leqslant 14$ &19.3941\\ \hline

      $7p(1|1,1) +3p(1|1,2) -4p(1|1,3) +p(1|2,1) -7p(1|2,2) +6p(1|2,3) -4p(1|3,1) +4p(1|3,2) $&\\$+4p(1|3,3) -6p(1|4,1) -4p(1|4,2) -10p(1|4,3) \leqslant 10$ &13.7269\\ \hline

      $11p(1|1,1) +p(1|1,2) -10p(1|1,3) +p(1|2,1) -4p(1|2,2) +5p(1|2,3) -4p(1|3,1) +4p(1|3,2) $&\\$+7p(1|3,3) -13p(1|4,1) +p(1|4,2) -12p(1|4,3) \leqslant 12$ &16.296 \\ \hline
      
      $5p(1|1,1) +p(1|1,2) -4p(1|1,3) +p(1|2,1) -3p(1|2,2) -p(1|2,3) -2p(1|3,1) +2p(1|3,2) $&\\$+4p(1|3,3) -6p(1|4,1) +3p(1|4,2) -9p(1|4,3) \leqslant 6$ & 7.9666 \\ \hline

      $6p(1|1,1) +4p(1|1,2) -4p(1|1,3) +2p(1|2,1) -2p(1|2,2) +4p(1|2,3) -6p(1|31) -3p(1|3,2) $&\\$ -9p(1|3,3) -8p(1|4,1) +3p(1|4,2) +5p(1|4,3) \leqslant 10$ & 12.8339\\ \hline

      $11p(1|1,1) +7p(1|1,2) +4p(1|1,3) +9p(1|2,1) -11p(1|2,2) -2p(1|2,3) -3p(1|3,1) +3p(1|3,2) $&\\$ -4p(1|3,3) -9p(1|4,1) -3p(1|4,2) +6p(1|4,3) \leqslant 22$ & 26.0756 \\ \hline

      $\mathcal{I}_5=5p(1|1,1) +3p(1|1,2) -2p(1|1,3) +2p(1|2,1) -5p(1|2,2) +3p(1|2,3) -5p(1|3,1) +2p(1|3,2) $&\\$ +5p(1|3,3) -6p(1|4,1) -2p(1|4,2) -8p(1|4,3) \leqslant 8$ & 10.8596 \\ \hline

      $11p(1|1,1) +p(1|1,2) -10p(1|1,3) +3p(1|2,1) -3p(1|2,2) +6p(1|2,3) -3p(1|3,1) +5p(1|3,2) $&\\$ +2p(1|3,3) -7p(1|4,1) -p(1|4,2) -6p(1|4,3) \leqslant 14$ & 17.2020 \\ \hline
      
      $5p(1|1,1) -2p(1|1,2) -7p(1|1,3) +p(1|2,1) +2p(1|2,2) +3p(1|2,3) -3p(1|3,1) -2p(1|3,2) $&\\$+3p(1|3,3) -7p(1|4,1) +4p(1|4,2) -9p(1|4,3) \leqslant 6$ & 7.5839 \\ \hline

      $5p(1|1,1) +p(1|1,2) -4p(1|1,3) +p(1|2,1) -p(1|2,2) +2p(1|2,3) -3p(1|3,1) -3p(1|3,2) $&\\$-7p(1|4,1) +5p(1|4,2) -2p(1|4,3) \leqslant 6$ & 6.9577 \\ \hline

      $11p(1|1,1) +p(1|1,2) -10p(1|1,3) +7p(1|2,1) -p(1|2,2) +8p(1|2,3) -3p(1|3,1) -3p(1|3,2) $&\\$ -7p(1|4,1) +5p(1|4,2) -2p(1|4,3) \leqslant 18$ & 21.20202 \\ \hline

      $6p(1|1,1) -p(1|1,2) -7p(1|1,3) +4p(1|2,1) +p(1|2,2) +5p(1|2,3) -2p(1|3,1) -4p(1|3,2) $&\\$ +2p(1|3,3) -4p(1|4,1) +2p(1|4,2) -2p(1|4,3) \leqslant 10$ &  12.0199 \\ \hline

      $11p(1|1,1) +9p(1|1,2) -2p(1|1,3) +6p(1|2,1) -10p(1|2,2) -p(1|3,1) +p(1|3,2) +2p(1|3,3) $&\\$-6p(1|4,1) +4p(1|4,2) -4p(1|4,3) \leqslant 20$ & 23.3969 \\ \hline

      $4p(1|1,1) +3p(1|1,2) -2p(1|1,3) +3p(1|2,1) -5p(1|2,2) -3p(1|2,3) -p(1|3,1) +p(1|3,2) $&\\$+2p(1|3,3) -3p(1|4,1) +3p(1|4,2) -6p(1|4,3) \leqslant 7$ & 7.8917\\ \hline
      

      $* 6p(1|1,1) -2p(1|1,2) -6p(1|1,3) +p(1|2,1) +2p(1|2,2) -p(1|2,3) -2p(1|3,1) +2p(1|3,2) $&\\$+4p(1|3,3) -7p(1|4,1) +3p(1|4,2) -9p(1|4,3) \leqslant 7$ & 9.0713 \\ \hline

      $10p(1|1,1) -7p(1|1,3) +p(1|2,1) -6p(1|2,2) -4p(1|2,3) -2p(1|3,1) +3p(1|3,2) +5p(1|3,3) $&\\$-8p(1|4,1) +8p(1|4,2) -16p(1|4,3) \leqslant 11$ & 11\\ \hline

      $5p(1|1,1) +p(1|1,2) -5p(1|1,3) +p(1|2,1) -2p(1|2,2) -p(1|2,3) -2p(1|3,1) +2p(1|3,2)$&\\$ +4p(1|3,3) -6p(1|4,1) +3p(1|4,2) -9p(1|4,3) \leqslant 6$ &  7.8464 \\ \hline

      $10p(1|1,1) -p(1|1,2) -11p(1|1,3) +6p(1|2,1) +p(1|2,2) +7p(1|2,3) -2p(1|3,1) -5p(1|3,2) $&\\$+3p(1|3,3) -6p(1|4,1) +3p(1|4,2) -3p(1|4,3) \leqslant 16$ & 19.2020 \\ \hline

    \end{tabular}
    
    \label{tab:placeholder}
\end{table}

\begin{table}[]
    \centering
    \begin{tabular}{|c|c|}
    \hline
    Inequalities &  $\mathcal{S_Q}_2^L$  \\ \hline

      $5p(1|1,1) -5p(1|1,3) +p(1|2,1) -2p(1|2,2) -p(1|2,3) -2p(1|3,1) +2p(1|3,2) +4p(1|3,3) $&\\$ -6p(1|4,1) +4p(1|4,2) -10p(1|4,3) \leqslant 6$ & 7.7519\\ \hline
    
      $7p(1|1,1) +5p(1|1,2) -2p(1|1,3) +4p(1|2,1) -6p(1|2,2) -2p(1|2,3) -p(1|3,1) +p(1|3,2) $&\\$ +2p(1|3,3) -4p(1|4,1) +4p(1|4,2) -6p(1|4,3) \leqslant 12$ & 13.9010\\ \hline
      
      $4p(1|1,1) -4p(1|1,3) +p(1|2,1) -2p(1|2,2) +2p(1|2,3) -p(1|3,1) +2p(1|3,2) +3p(1|3,3) $&\\$ -5p(1|4,1) -5p(1|4,3) \leqslant 5$ & 6.6008 \\ \hline
    
      $3p(1|1,1) -3p(1|1,3) -p(1|2,1) +p(1|2,2) +2p(1|2,3) -p(1|3,1) -2p(1|3,2) +p(1|3,3) $&\\$ -5p(1|4,1) +2p(1|4,2) -4p(1|4,3) \leqslant 3$ & 3.7468\\ \hline

      $5p(1|1,1) -2p(1|1,2) -4p(1|1,3) -3p(1|2,2) +3p(1|2,3) -p(1|3,1) +3p(1|3,2) +2p(1|3,3) $&\\$ -5p(1|4,1) -2p(1|4,2) -3p(1|4,3) \leqslant 5$ &  6.5951\\ \hline

      $9p(1|1,1) +7p(1|1,2) -2p(1|1,3) +5p(1|2,1) -8p(1|2,2) +p(1|2,3) -p(1|3,1) +p(1|3,2)$&\\$ +2p(1|3,3) -5p(1|4,1) +2p(1|4,2) -3p(1|4,3) \leqslant 16 $ & 18.8316\\ \hline

      $3p(1|1,1) -p(1|1,2) -4p(1|1,3) +p(1|2,1) +p(1|2,2) +2p(1|2,3) -3p(1|3,1) -5p(1|3,2) $&\\$+2p(1|3,3) -5p(1|4,1) +3p(1|4,2) -2p(1|4,3) \leqslant 4$ & 5.2611 \\ \hline

      $4p(1|1,1) -p(1|1,3) -4p(1|2,2) -4p(1|2,3) -p(1|3,1) +p(1|3,2) +p(1|3,3) -2p(1|4,1) $&\\$ -5p(1|4,2) +3p(1|4,3) \leqslant 4$ & 5.2699\\ \hline

      $4p(1|1,1) -2p(1|1,2) -5p(1|1,3) +p(1|2,1) +2p(1|2,2) -p(1|2,3) -p(1|3,1) +p(1|3,2) $&\\$+3p(1|3,3) -5p(1|4,1) +2p(1|4,2) -6p(1|4,3) \leqslant 5$ &  6.5269 \\ \hline

      $4p(1|1,1) +2p(1|1,2) -2p(1|1,3) -2p(1|2,2) +2p(1|2,3) -4p(1|3,1) +3p(1|3,2) +3p(1|3,3) $&\\$ -4p(1|4,1) -p(1|4,2) -5p(1|4,3) \leqslant 6$ &  6.4866\\ \hline

      $8p(1|1,1) -5p(1|1,3) +p(1|2,1) -5p(1|2,2) -4p(1|2,3) -2p(1|3,1) +2p(1|32) +4p(1|3,3) $&\\$ -6p(1|4,1) +7p(1|4,2) -13p(1|4,3) \leqslant 9$ & 9 \\ \hline

      $3p(1|1,1) -2p(1|1,2) -3p(1|1,3) +p(1|2,1) +2p(1|2,2) +p(1|2,3) -p(1|3,1) +p(1|3,3) $&\\$-3p(1|4,1) +2p(1|4,2) -5p(1|4,3) \leqslant 4$ & 5.3289\\ \hline

      $7p(1|1,1) +5p(1|1,2) -2p(1|1,3) +3p(1|2,1) -5p(1|2,2) -2p(1|2,3) -3p(1|3,1) +3p(1|3,2) $&\\$+2p(1|3,3) -3p(1|4,1) +3p(1|4,2) -6p(1|4,3) \leqslant 12$ & 13.5847\\ \hline

      $4p(1|1,1) -p(1|1,2) -5p(1|1,3) +2p(1|2,1) +p(1|2,2) +3p(1|2,3) -p(1|3,1) -3p(1|3,2) $&\\$+2p(1|3,3) -3p(1|4,1) +p(1|4,2) -2p(1|4,3) \leqslant 6$ & 7.7333\\ \hline

      $6p(1|1,1) +6p(1|1,2) +4p(1|2,1) -5p(1|2,2) -p(1|2,3) -p(1|3,1) -p(1|3,2) +p(1|3,3) $&\\$ -3p(1|4,1) +3p(1|4,2) -3p(1|4,3) \leqslant 12$ & 14.0702 \\ \hline

      $5p(1|1,1) -3p(1|1,2) -5p(1|1,3) +p(1|2,1) +3p(1|2,2) +2p(1|2,3) -p(1|3,1) -4p(1|3,2) $&\\$ +4p(1|3,3) -4p(1|4,1) -4p(1|4,3) \leqslant 6$ & 7.9663\\ \hline

      $8p(1|1,1) -5p(1|1,2) -7p(1|1,3) +2p(1|2,1) +5p(1|2,2) +3p(1|2,3) -2p(1|3,1) -p(1|3,2) $&\\$ +p(1|3,3) -6p(1|4,1) +5p(1|4,2) -11p(1|4,3) \leqslant 10$ &  12.5868\\ \hline

      $10p(1|1,1) +5p(1|1,2) -5p(1|1,3) +5p(1|2,1) -7p(1|2,2) +p(1|2,3) +2p(1|3,1) +3p(1|3,3) $&\\$-3p(1|4,1) +p(1|4,2) -2p(1|4,3) \leqslant 17$ & 18.9604\\ \hline

      $16p(1|1,1) -5p(1|1,2) -15p(1|1,3) -2p(1|2,1) +7p(1|2,2) +9p(1|2,3) -2p(1|3,1) -3p(1|3,2) $&\\$+p(1|3,3) -14p(1|4,1) +9p(1|4,2) -23p(1|4,3) \leqslant 16$ & 16\\ \hline

      $2p(1|1,1) +2p(1|1,2) -2p(1|1,3) +p(1|2,1) -p(1|2,2) +2p(1|2,3) -2p(1|3,1) -2p(1|3,2) $&\\$-4p(1|3,3) -3p(1|4,1) +p(1|4,2) +2p(1|4,3) \leqslant 4$ &  5.0191 \\ \hline

      $2p(1|1,1) -2p(1|1,3) +p(1|2,1) -2p(1|2,2) -p(1|2,3) +p(1|3,2) +p(1|3,3) -2p(1|4,1)$&\\$ +2p(1|4,2) -4p(1|4,3) \leqslant 3$ & 3\\ \hline

      $3p(1|1,1) +3p(1|1,2) -2p(1|1,3) +2p(1|2,1) -2p(1|2,2) -4p(1|2,3) +p(1|3,1) -p(1|3,2) $&\\$ +2p(1|3,3) -4p(1|4,1) +2p(1|4,2) -2p(1|4,3) \leqslant 6$ &  6.7692\\
      \hline
    \end{tabular}
    
    \label{tab:placeholder}
\end{table}

\begin{table}[h]
    \centering
    \begin{tabular}{|c|c|c|}
    \hline
    Inequalities &  $\mathcal{S_Q}_2^L$ \\ \hline

    $3p(1|1,1) +2p(1|1,2) -p(1|1,3) +2p(1|2,1) -3p(1|2,2) -2p(1|2,3) -p(1|3,1) +p(1|3,2)$&\\$ +p(1|3,3) -2p(1|4,1) +2p(1|4,2) -4p(1|4,3) \leqslant 5$ &  5.8829\\ \hline

$4p(1|1,1) -p(1|1,2) -4p(1|1,3) -p(1|2,1) +p(1|2,2) +3p(1|2,3) -2p(1|3,1) +3p(1|3,2)$&\\$ -3p(1|3,3) -4p(1|4,1) -p(1|4,2) -3p(1|4,3) \leqslant 4$  & 4\\ \hline

$5p(1|1,1) -p(1|1,2) -4p(1|1,3) +p(1|2,1) -2p(1|2,2) +3p(1|2,3) -p(1|3,1) +3p(1|3,2)$&\\$ +2p(1|3,3) -5p(1|4,1) -2p(1|4,2) -3p(1|4,3) \leqslant 6$ &8.3323 \\ \hline

$5p(1|1,1) +3p(1|1,2) -2p(1|1,3) +3p(1|2,1) -3p(1|2,2) +3p(1|2,3) +p(1|3,1) -4p(1|3,2) $&\\$-3p(1|3,3) -p(1|4,1) +2p(1|4,2) -2p(1|4,3) \leqslant 9$ & 10.5274 \\ \hline

$4p(1|1,1) -p(1|1,2) -5p(1|1,3) +2p(1|2,1) +3p(1|2,2) +p(1|2,3) +2p(1|3,1) -3p(1|3,2)$&\\$ +3p(1|3,3) -4p(1|4,1) +p(1|4,2) -3p(1|4,3) \leqslant 8$ & 9.3277\\ \hline

$6p(1|1,1) +2p(1|1,2) -4p(1|1,3) +5p(1|2,1) -2p(1|2,2) +3p(1|2,3) -3p(1|3,2) +p(1|3,3)$&\\$ -p(1|4,1) -p(1|4,3) \leqslant 11$ &  12.5586\\ \hline

$2p(1|1,1) +p(1|1,2) -p(1|1,3) +p(1|2,1) -2p(1|2,2) +p(1|2,3) -2p(1|3,1) +p(1|3,2)$&\\$ +2p(1|3,3) -3p(1|4,1) -p(1|4,2) -3p(1|4,3) \leqslant 3$ & 4.01  \\ \hline


$3p(1|1,1) -p(1|1,2) -3p(1|1,3) +p(1|2,2) -p(1|2,3) -p(1|3,1) +p(1|3,2) +2p(1|3,3)$&\\$ -3p(1|4,1) +p(1|4,2) -4p(1|4,3) \leqslant 3$ & 3.8467 \\ \hline

$3p(1|1,1) +2p(1|1,2) -p(1|1,3) +p(1|2,1) -2p(1|2,2) +2p(1|2,3) +p(1|3,1)$&\\$ -3p(1|3,2) -2p(1|3,3) -p(1|4,1) +p(1|4,2) -p(1|4,3) \leqslant 5$ & 5.9207  \\ \hline

$3p(1|1,1) +2p(1|1,2) -p(1|1,3) +2p(1|2,1) -3p(1|2,2) +p(1|2,3) -p(1|3,1) +p(1|3,2)$&\\$ +p(1|3,3) -2p(1|4,1) -p(1|4,2) -2p(1|4,3) \leqslant 5$&  6.1056 \\ \hline
    \end{tabular}
    \caption{Facets of $(4,3,2)$ dimension bound scenario and their quantum violations are enlisted.}
    \label{table3}
\end{table}
\newpage
\section{Distinguishability bound scenario}
\subsection{$(3,3,2)$ scenario}
In this scenario we get $1073741824$ number of vertices, from this we get facet inequalitites. We enlisted some of them. Each facet inequality shows quantum advantage over the classical one i.e, $\mathcal{S_Q} > \mathcal{S_C}$. We also enlist the quantum lower bound $\mathcal{S_Q}_2^L$ that is obtained from See-Saw technique. We discuss about upper bound $\mathcal{S_Q}$ using semidefinte hierarchy discussed in Section \ref{sdphierarchy}, for inequality $\mathcal{I}_6$. In general our hierarchy technique doesn't converge with the lower bound $\mathcal{S_Q}_2^L$, we check upto dimension $5$. There might be some higher dimension where this will converge.
\begin{table}[htpb]
     \centering
    \begin{tabular}{|c|c|c|}
    \hline
    Inequalities & $\mathcal{S_C}$& $\mathcal{S_Q}_2^L$  \\ \hline


     $ p(1|3,1) +p(1|1,2) -2p(1|2,2) +p(1|1,3) +p(1|2,3) -p(1|3,3) \leqslant 3D1 +3D2  -1$&3& 3 \\ \hline

    $p(1|3,1) +p(1|1,2) +p(1|2,2) -3p(1|3,2) +p(1|1,3) -p(1|2,3) \leqslant 6D1 + 3D2 -3$ & 3& 3.1936 \\ \hline

    $p(1|2,1) +p(1|3,1) +p(1|1,2) -4p(1|2,2) +2p(1|3,2) +2p(1|1,3) -2p(1|3,3)$&&\\$ \leqslant 9D1 +3D2  -3$ &5&5.1315\\ \hline

    $p(1|2,1) +p(1|3,1) +2p(1|1,2) -6p(1|2,2) +2p(1|3,2) +2p(1|1,3) +p(1|2,3) -3p(1|3,3)$&&\\$ \leqslant 12D1 + 6D2  -6$ & 6 &  6.4676\\ \hline


    $ p(1|2,1) +p(1|3,1) +3p(1|1,2) -3p(1|2,2) -p(1|3,2) +p(1|1,3) +3p(1|2,3) -4p(1|3,3)$&&\\$ \leqslant 12D1 +3D2  -4$ & 6& 6.2009\\ \hline

    $2p(1|2,1) +2p(1|3,1) +p(1|1,2) -3p(1|2,2) +2p(1|3,2) +p(1|1,3) +p(1|2,3) -6p(1|3,3)$&&\\$ \leqslant 6D1 +12D2  -6$&6 & 6.4676 \\ \hline

    $p(1|1,1) -2p(1|2,1) -2p(1|3,1) +p(1|1,2) -2p(1|2,2) +2p(1|3,2) +p(1|1,3) +4p(1|2,3) -p(1|3,3)$&&\\$ \leqslant 3D1 +9D2  -2$ &6&  6.0756 \\ \hline

    $p(1|1,1) -2p(1|2,1) +p(1|3,1) -p(1|2,2) -p(1|3,2) +p(1|1,3) +3p(1|2,3) -p(1|3,3)$&&\\$ \leqslant 3D1 +6D2 -2$& 4 & 4.3261\\ \hline

    $p(1|1,1) -2p(1|2,1) +3p(1|3,1) -p(1|1,2) +5p(1|2,2) -p(1|3,2) +p(1|1,3) -3p(1|2,3) -5p(1|3,3)$&&\\$ \leqslant 3D1 +15D2  -6$ & 6 & 6.5741 \\ \hline

    $\mathcal{I}_6=p(1|1,1) -p(1|2,1) -p(1|3,1) -2p(1|1,2) +2p(1|2,2) +p(1|3,2) +p(1|1,3) +3p(1|2,3) -2p(1|3,3)$&&\\$ \leqslant 6D1 +3D2  -1$ & $5$&5.5348 \\ \hline

    $p(1|1,1) -p(1|2,1) -p(1|3,1) -p(1|1,2) +p(1|2,2) +p(1|3,2) +p(1|1,3) +2p(1|2,3) -p(1|3,3)$&&\\$ \leqslant 3D1 +3D2 $ & 4& 4.1820\\ \hline

    $p(1|1,1) -p(1|2,1) +p(1|3,1) -3p(1|1,2) +3p(1|2,2) -p(1|3,2) +4p(1|1,3) +2p(1|2,3) -6p(1|3,3)$&&\\$ \leqslant 18D1 +3D2 -7$& 7&7.1448 \\ \hline

    $p(1|1,1) -p(1|2,1) +p(1|3,1) -2p(1|1,2) +3p(1|2,2) -p(1|3,2) +3p(1|1,3) +p(1|2,3) -5p(1|3,3)$&&\\$ \leqslant 15D1 +3D2  -6$ & 6 &6.1490 \\ \hline

    $p(1|1,1) -p(1|2,1) +p(1|3,1) -p(1|1,2) +3p(1|2,2) -2p(1|3,2) +2p(1|1,3) -p(1|2,3) -3p(1|3,3)$&&\\$ \leqslant 9D1 +3D2  -4$ & 4 & 4.4861 \\ \hline

    $p(1|1,1) -p(1|2,1) +p(1|3,1) +2p(1|1,2) -p(1|2,2) -3p(1|3,2) -p(1|1,3) +2p(1|2,3) -2p(1|3,3)$&&\\$ \leqslant 6D1 +3D2  -3$ & 3 & 3.5348\\ \hline

    $ p(1|1,1) -p(1|2,1) +2p(1|3,1) +2p(1|1,2) -2p(1|2,2) -4p(1|3,2) -p(1|1,3) +3p(1|2,3) -2p(1|3,3)$&&\\$ \leqslant 6D1 + 6D2 -4$ & 4 & 4.4259 \\ \hline

    $p(1|1,1) +p(1|3,1) -3p(1|1,2) +3p(1|2,2) -p(1|3,2) +3p(1|1,3) +p(1|2,3) -4p(1|3,3)$&&\\$ \leqslant 12D1 +3D2  -4$ & 6 & 6.2009\\ \hline

    $p(1|1,1) +2p(1|3,1) +3p(1|1,2) +p(1|2,2) -3p(1|3,2) +2p(1|1,3) -5p(1|2,3) +2p(1|3,3)$&&\\$ \leqslant 12D1 +3D2  -2$ &8 & 8.8264\\ \hline

    $p(1|1,1) +p(1|2,1) +p(1|3,1) -2p(1|1,2) -2p(1|2,2) +3p(1|3,2) +3p(1|1,3) -2p(1|2,3) -p(1|3,3)$&&\\$ \leqslant 9D1 +3D2  -2$ &6 & 6.0522\\ \hline

    $2p(1|1,1) -2p(1|2,1) -p(1|3,1) -p(1|1,2) +p(1|2,2) +p(1|3,2) +p(1|1,3) +3p(1|2,3) -2p(1|3,3)$&&\\$ \leqslant 6D1 +3D2  -1$ &5 & 5.1820 \\ \hline

    $2p(1|1,1) -2p(1|2,1) +2p(1|3,1) -p(1|1,2) +4p(1|2,2) -3p(1|3,2) +3p(1|1,3) -2p(1|2,3) -5p(1|3,3)$&&\\$ \leqslant 12D1 +6D2 -6$&6 & 6.7506\\ \hline

    $2p(1|1,1) -p(1|2,1) -p(1|3,1) -2p(1|1,2) +5p(1|2,2) -4p(1|3,2) -2p(1|1,3) +p(1|2,3) +3p(1|3,3)$&&\\$ \leqslant 15D1 + 3D2 -5$& 7 & 7.0770 \\ \hline
$2p(1|1,1) -p(1|2,1) +p(1|3,1) -2p(1|1,2) +4p(1|2,2) -2p(1|3,2) -4p(1|1,3) -p(1|2,3) +3p(1|3,3)$&&\\$ \leqslant 12D1 +3D2 -4$ & 6 & 6.3640\\ \hline
    
    \end{tabular}
    \label{tab:placeholder}
\end{table}

\begin{table}[]
    \centering
    \begin{tabular}{|c|c|c|c|}
    \hline
    Inequalities & $\mathcal{S_C}$& $\mathcal{S_Q}_2^L$ \\ \hline

    $2p(1|1,1) -p(1|2,1) +2p(1|3,1) +5p(1|1,2) -3p(1|2,2) -5p(1|3,2) -p(1|1,3) +4p(1|2,3) -3p(1|3,3)$&&\\$ \leqslant 15D1 +6D2  -6$&8&8.7425 \\ \hline

    $2p(1|1,1) +4p(1|2,1) -2p(1|3,1) -2p(1|1,2) -3p(1|2,2) +5p(1|3,2) +4p(1|1,3) -7p(1|2,3) -3p(1|3,3)$&&\\$ \leqslant 12D1 +12D2  -8$ & 8 & 8.5456 \\ \hline

    $3p(1|1,1) -4p(1|2,1) -p(1|3,1) +p(1|1,2) +6p(1|2,2) -p(1|3,2) -2p(1|1,3) -2p(1|2,3) +p(1|3,3)$&&\\$ \leqslant 3D1 +15D2 -5$ & 7&  7.1589 \\ \hline

    $3p(1|1,1) -4p(1|2,1) +4p(1|3,1) +p(1|1,2) +p(1|2,2) -10p(1|3,2) -p(1|1,3) +3p(1|2,3) +4p(1|3,3)$&&\\$ \leqslant 6D1 +24D2 -9$ &11 &  11.6304\\ \hline

    $3p(1|1,1) -3p(1|2,1) +3p(1|3,1) -2p(1|1,2) +6p(1|2,2) -4p(1|3,2) +5p(1|1,3) -2p(1|2,3) -7p(1|3,3)$&&\\$ \leqslant 21D1 +9D2 -10$ & 10 & 10.8852\\ \hline

    $3p(1|1,1) -2p(1|2,1) -4p(1|3,1) -p(1|1,2) +4p(1|2,2) -p(1|3,2) +p(1|1,3) -2p(1|2,3) +4p(1|3,3)$&&\\$ \leqslant 3D1 +15D2 -4$ & 8 &8\\ \hline

    $3p(1|1,1) -2p(1|2,1) +p(1|3,1) -3p(1|1,2) +2p(1|2,2) +p(1|3,2) +4p(1|1,3) +4p(1|2,3) -10p(1|3,3)$&&\\$ \leqslant 24D1 +6D2  -10$& 10 &10.3271 \\ \hline

    $3p(1|1,1) -p(1|2,1) -2p(1|3,1) -2p(1|1,2) +2p(1|2,2) -4p(1|3,2) -p(1|1,3) +p(1|2,3) +2p(1|3,3)$&&\\$ \leqslant 6D1 +6D2 -4$ & 4 & 4.4259\\ \hline

    $3p(1|1,1) -p(1|2,1) -p(1|3,1) +p(1|1,2) -3p(1|2,2) -p(1|3,2) -2p(1|1,3) -3p(1|2,3) +5p(1|3,3)$&&\\$ \leqslant 15D1 +3D2 -6$&6 & 6.5741 \\ \hline

    $3p(1|1,1) -p(1|2,1) -p(1|3,1) +2p(1|1,2) -4p(1|2,2) +2p(1|3,2) -3p(1|1,3) -p(1|2,3) +3p(1|3,3)$&&\\$ \leqslant 18D1 +3D2 -7$ & 7 &  7.1154\\ \hline

    $3p(1|1,1) -p(1|2,1) +p(1|3,1) +3p(1|1,2) +p(1|2,2) -5p(1|3,2) +p(1|1,3) -3p(1|2,3) +2p(1|3,3)$&&\\$ \leqslant 15D1 +3D2 -4$&8 &  8.6007 \\ \hline

    $3p(1|1,1) -p(1|2,1) +p(1|3,1) +4p(1|1,2) +4p(1|2,2) -10p(1|3,2) -4p(1|1,3) +3p(1|2,3) +p(1|3,3)$&&\\$ \leqslant 24D1 +6D2 -9$ & 11 & 11.6304 \\ \hline
    
    $3p(1|1,1) +p(1|2,1) -2p(1|3,1) -5p(1|1,2) +3p(1|2,2) +p(1|3,2) +p(1|1,3) -2p(1|2,3) +3p(1|3,3)$&&\\$ \leqslant 12D1 + 3D2 -2$ & 8  & 8.0836 \\ \hline
    
    $3p(1|1,1) +5p(1|2,1) -2p(1|3,1) -3p(1|1,2) -2p(1|2,2) +6p(1|3,2) +3p(1|1,3) -7p(1|2,3) -4p(1|3,3)$&&\\$ \leqslant 9D1 +21D2 -10$ &10 &  10.8852 \\ \hline

    $4p(1|1,1) -4p(1|2,1) +4p(1|3,1) -2p(1|1,2) +7p(1|2,2) -5p(1|3,2) +6p(1|1,3) -3p(1|2,3) -9p(1|3,3)$&&\\$ \leqslant 24D1 +12D2 -12$&12 &  13.1542  \\ \hline

    $4p(1|1,1) -4p(1|2,1) +3p(1|3,1) -p(1|1,2) +6p(1|2,2) -4p(1|3,2) +5p(1|1,3) -2p(1|2,3) -7p(1|3,3)$&&\\$ \leqslant 21D1 +9D2 -9$ &11 & 11.8679 \\ \hline

    $4p(1|1,1) -2p(1|2,1) +2p(1|3,1) -9p(1|1,2) +2p(1|2,2) +7p(1|3,2) -p(1|1,3) +4p(1|2,3) -5p(1|3,3)$&&\\$ \leqslant 24D1 +6D2  -8$&12 &12.2275 \\ \hline
    
    $5p(1|1,1) -2p(1|2,1) -p(1|3,1) -3p(1|1,2) +2p(1|2,2) +5p(1|3,2) -2p(1|1,3) +4p(1|2,3) -6p(1|3,3)$&&\\$ \leqslant 6D1 +18D2  -6$& 10 & 10.2466 \\ \hline

    $7p(1|1,1) -5p(1|2,1) +2p(1|3,1) -p(1|1,2) +3p(1|2,2) -2p(1|3,2) -8p(1|1,3) -4p(1|2,3) +10p(1|3,3)$&&\\$ \leqslant 30D1 +6D2  -10$&14 & 14.1540 \\ \hline

    $8p(1|1,1) -6p(1|2,1) +2p(1|3,1) -7p(1|1,2) -3p(1|2,2) +4p(1|3,2) +p(1|1,3) +3p(1|2,3) -2p(1|3,3)$&&\\$ \leqslant 12D1 +12D2  -6$&10 & 10.5524  \\ \hline

    $8p(1|1,1) -5p(1|2,1) +2p(1|3,1) -9p(1|1,2) -4p(1|2,2) +13p(1|3,2) -2p(1|1,3) +p(1|2,3) -3p(1|3,3)$&&\\$ \leqslant 30D1 +9D2  -11$&15 & 15.8504 \\ \hline

    $9p(1|1,1) -7p(1|2,1) +2p(1|3,1) -11p(1|1,2) -7p(1|2,2) +15p(1|3,2) -2p(1|1,3) +4p(1|2,3)$&&\\$ -2p(1|3,3) \leqslant 45D1 +9D2  -17$ & 19 &  19.2028 \\ \hline

    \end{tabular}
\end{table}

\begin{table}[]
    \centering
    \begin{tabular}{|c|c|c|}
    \hline
    Inequalities & $\mathcal{S_C}$& $\mathcal{S_Q}_2^L$  \\ \hline

    $2p(1|1,1) +p(1|2,1) +p(1|3,1) -4p(1|1,2) +3p(1|2,2) +p(1|3,2) +3p(1|1,3) +2p(1|2,3) -7p(1|3,3)$&&\\$ \leqslant 15D1 +6D2  -5$ & 9 & 9.6970\\ \hline
    
    $9p(1|1,1) -6p(1|2,1) +3p(1|3,1) -11p(1|1,2) -4p(1|2,2) +15p(1|3,2) -2p(1|1,3) +2p(1|2,3) -4p(1|3,3)$&&\\$ \leqslant 36D1 +9D2  -12$ & 18 & 18.7450 \\ \hline

    $15p(1|1,1) +4p(1|2,1) -10p(1|3,1) -5p(1|1,2) +3p(1|2,2) -2p(1|3,2) +9p(1|1,3) -17p(1|2,3) +8p(1|3,3)$&&\\$ \leqslant 48D1 +12D2  -15$ & 25 & 26.7253 \\ \hline

    $17p(1|1,1) -9p(1|2,1) -8p(1|3,1) -2p(1|1,2) -11p(1|2,2) +9p(1|3,2) -p(1|1,3) +7p(1|2,3) +p(1|3,3)$&&\\$ \leqslant 48D1 +12D2  -17$& 23& 23.0408  \\ \hline

    $28p(1|1,1) -5p(1|2,1) -3p(1|3,1) -12p(1|1,2) -2p(1|2,2) +10p(1|3,2) -10p(1|1,3) +7p(1|2,3) $&&\\$ -13p(1|3,3) \leqslant 18D1 +72D2  -30$ & 30 & 30.1046 \\ \hline

$-p(1|1,1) -2p(1|2,1) +p(1|3,1) +p(1|1,2) +2p(1|2,2) -3p(1|3,2) -2p(1|1,3) +4p(1|2,3) +2p(1|3,3)$&&\\$ \leqslant 6D1 +6D2  -2$ & 6 & 6.4259\\ \hline

    $p(1|1,1) -12p(1|2,1) -5p(1|3,1) -p(1|1,2) +12p(1|2,2) -11p(1|3,2) -2p(1|1,3) +4p(1|2,3) +13p(1|3,3)$&&\\$ \leqslant 12D1 +45D2  -19$ & 19& 20.1960  \\ \hline

    $ p(1|1,1) -6p(1|2,1) +3p(1|3,1) +p(1|1,2) +4p(1|2,2) -14p(1|3,2) -2p(1|1,3) +6p(1|2,3) +6p(1|3,3)$&&\\$ \leqslant 12D1 +30D2  -14$ & 14 & 14.3004\\ \hline

    $8p(1|1,1) -8p(1|2,1) -6p(1|3,1) -3p(1|1,2) +13p(1|2,2) -16p(1|3,2) -5p(1|1,3) +5p(1|2,3) +10p(1|3,3)$&&\\$ \leqslant 30D1 +30D2  -20$ & 20 & 20.6601\\ \hline

    \end{tabular}
    \caption{Facets of (3,3,2)-scenario bounded by distinguishability. Here $\mathcal{S_C}$, $\mathcal{S_Q}^L_2$, corresponding to the $D_1=D_2=2/3$. $\mathcal{S_Q}^L_2$ is the lower bound in $d=2$ that we get from SeeSaw optimization respectively.}
    \label{table4}
\end{table}

\newpage
\section{Semi-definite optimization to obtain upper bound}\label{hierarchy}
We check the upper bound for the multiparty communication scenario bounded by both distinguishability and dimension constraints. Building on the approach for bounding the distinguishability constraints scenario, as discussed in the \cite{pandit2025limitsclassicalcorrelationsquantum}, we have to introduce two auxiliary variables in order to bounding the distinguishability of the sender's input in a multiparty communication scenario. Here we present the hierarchy of semidefinite programming for the set $\mathcal{Q}$, which falls under commutative polynomial optimization. The idea was first introduced by \cite{Tavakoli2022informationally} in order to bound the informational restricted correlation. Later in another approach \cite{PhysRevResearch.6.043269}, which is very computationally easy to implement, takes less amount of time to get the convergence of the optimal solution, and get the same result $\mathcal{S_Q}$ for bounding the distinguishability in the sender's input. Both approaches consists of only a single sender and single receiver communication task, hence it is a noncommutative polynomial optimization. We here present a new hierarchy approach for bounding the distinguishability of two senders and a single receiver communication task, where the receiver has no input, implying the receiver always has a fixed input in measurement. For any positive semidefinite operator $\mathcal{O} \geqslant0$.  Without loss of generality, we consider the pure states and projective measurement to construct $\Gamma$. To find the upper bound, we consider a list of operators $\mathcal{O}=\{ \mathbb{I}_{x} \otimes \mathbb{I}_{y}, \{\Theta \otimes \mathbb{I}_{y}\}, \{\mathbb{I}_{x} \otimes \Phi \},  \{\rho_{x} \otimes \mathbb{I}_{y}\}_{x}, \{\mathbb{I}_{x} \otimes \sigma_{y} \}_{y}, \{ M_{z} \}_{z} \}$. From $\mathcal{O}$, we construct monomial list $\mathcal{L}= \{ \mathbb{I}_{x} \otimes \mathbb{I}_{y}, \{\rho_{x} \otimes \mathbb{I}_{y} \}_{y}, \{\mathbb{I}_{x} \otimes \sigma_{y} \}_{y} , \{ M_{z} \}_{z} , \{(\rho_{x} \otimes \mathbb{I}_{y})M_z \}_{x,z} \}, \{(\mathbb{I} \otimes \sigma_{y})M_z \}_{y,z} \} $. Here by default we fix the receiver's input y for all time. From these monomials, we build the moment matrix $\Gamma \geqslant 0$ (positive semidefinite), with elements $\Gamma_{u,v}=\Tr[uv\dagger]$, where $u,v \in [\mathcal{L}]$. This optimization is tracial commutative optimization problem, where tensor product is replaced by the commutation relation between the sender's states. We also consider additional states of auxiliary variables $\Theta$ and $\Phi$, to bound the distinguishability of both Alice and Bob  states respectively. So bounding the distinguishability of two senders implies,

\bea
\Gamma_{\Theta} \geqslant \Gamma_{\rho_{x}}, ~\Gamma_{\Phi} \geqslant \Gamma_{\sigma_{y}}, ~\frac{1}{n_{x}}\Gamma_{\Theta}\leqslant D_1, ~\frac{1}{n_{y}}\Gamma_{\Phi}\leqslant D_2.
\eea
Although we do not impose any restrictions on the dimension of the operators in the list $\mathcal{O}$, such that $\Gamma_{1,1}=\Tr[ \mathbb{I} \otimes \mathbb{I} ]$ is left as it is and $\Gamma_{\rho_{x},\sigma_{y},M_{z}} = \Tr[(\rho_x \otimes \sigma_y)M_z]$. Our semidifinite hierarchy techniques are valid for any dimension that we can. We compose the restriction of the entries of $\Gamma$ is
$\Tr[\rho_{x} \otimes \sigma_{y}]=1$(purity of states), orthogonality of the projective measurements is $M_{z} M_{z'}= \delta_{z,z'} M_{z}$ and bound on the distinguishability of states. Some entries of $\Gamma$ are correspond to the probabilities $p(z|x,y)=\Gamma_{\rho_{x},\sigma_{y},M_{z}}=\Tr[(\rho_{x} \otimes \sigma_{y})M_{z}]$. Combining all these, the hierarchy optimization rules are as follows:\\
\begin{align*} 
    \mathcal{S}_{\mathcal{Q}^2}^{max} =\max_{\{\rho_{x}\}^{n_{x}}_{x=1},\{\sigma_{y}\}^{n_{y}}_{y=1}, \Theta, \Phi, \{M_{z}\}} &\sum_{x,y,z} c_{x,y,z} p(z|x,y),\nonumber \\
     \text{such that}, \Gamma &\geqslant 0, \\ \nonumber
     x \in [n_{x}],  y \in [n_{y}],~\Gamma_{\rho_{x},\sigma_{y},M_{z}} &=\Gamma_{\sigma_{y},\rho_{x},M_{z}}, \\ \nonumber
     \sum_z \Gamma_{\rho_{x},\sigma_{y},M_{z}} &=\Gamma_{\rho_{x},\sigma_{y},\mathbb{I}}.~~~~~~~~~~~~~~~~~~  \nonumber \\
     \sum_z \Gamma_{\rho_{x},\mathbb{I},M_{z}} &=\Gamma_{\rho_{x},\mathbb{I},\mathbb{I}}.~~~~~~~~~~~~~~~~~~  \nonumber \\
    \sum_z \Gamma_{\mathbb{I},\sigma_{y},M_{z}} &=\Gamma_{\mathbb{I},\sigma_{y},\mathbb{I}}.~~~~~~~~~~~~~~~~~~  \nonumber \\
    \Gamma_{\Theta} \geqslant \Gamma_{\rho_{x}}, \nonumber 
    ~~~\frac{1}{n_{x}}\Gamma_{\Theta} &\leqslant D_1, \nonumber\\
    \Gamma_{\Phi} \geqslant \Gamma_{\sigma_{y}}, \nonumber 
    ~~~\frac{1}{n_{y}}\Gamma_{\Phi} &\leqslant D_2, \nonumber\\ \nonumber
    z\in\{0,\cdots, D-1\}, p(z|x,y) &=\Gamma_{\rho_{x},\sigma_{y},M_{z}}, \\ \nonumber
    \sum_{z=0}^{D-1} p(z|x,y) &=1, \\ \nonumber
    \Gamma_{\rho_{x},\sigma_{y}} &==1, \\ \nonumber
    \Gamma_{M_{z},M_{z'}} &= \delta_{z,z'} \Gamma_{\mathbb{I},M_{z}} \nonumber\\ 
    \nonumber 
\end{align*}

This semidefinite hierarchy corresponding to the multiparty communication bounded by the distinguishability of the sender's inputs. For the communication under dimension bounded scenario one simply cancel the distinguishability bound and add the dimension bound with $\Gamma_{\mathbb{I},\mathbb{I}} = \Tr(\mathbb{I} \otimes \mathbb{I}) = d^2$, $\Gamma_{\rho_x,\mathbb{I}} = \Tr(\rho_x \otimes \mathbb{I}) = d$, $\Gamma_{\mathbb{I},\sigma_y} = \Tr(\mathbb{I} \otimes \sigma_y) = d$, where $d$ is the dimension of the communication system.


\end{document}